\input{aipcheck}

\documentclass[
    ,final            
  ]
  {aipproc}

\layoutstyle{6x9}

\begin{document}

\title{Status Report of NNLO QCD Calculations}

\classification{12.38.Bx}
\keywords      {Perturbative QCD calculations}

\author{Michael Klasen\footnote{Permanent address:
  Laboratoire de Physique Subatomique et de Cosmologie,
  Universit\'e Joseph Fourier/CNRS-IN2P3, 53 Avenue des Martyrs,
  38026 Grenoble, France}}{address={Institute for Nuclear Theory,
  University of Washington, Box 351550, Seattle, WA 98195-1550, USA}
}

\begin{abstract}
 We review recent progress in next-to-next-to-leading order (NNLO)
 perturbative QCD calculations with special emphasis on results ready for
 phenomenological applications. Important examples are new results on
 structure functions and jet or Higgs boson production. In addition, we
 describe new calculational techniques based on twistors and their potential
 for efficient calculations of multiparticle amplitudes.
\end{abstract}

\maketitle

\vspace*{-9cm}
\noindent LPSC 05-054 \\
INT-ACK 05-038 \\
hep-ph/0506252
\vspace*{ 7cm}


\section{Introduction}

Despite the fact that the theory of strong interactions, quantum
chromodynamics (QCD), is today a well-established part of the Standard Model
of particle physics, it continues to be an extremely active field of
experimental and theoretical research. This is as true for non-perturbative
(and often lattice) determinations of meson and baryon spectra, decay
constants, or form factors for electric dipole moments or $B$-meson decays,
as it is for perturbative calculations. Today, the experimental precision of
high-energy collider data always requires calculations beyond the leading
order (LO) in the strong coupling constant, $\alpha_s(\mu)$, and often even
beyond the next-to-leading order (NLO) for reliable comparisons and
determinations of unknown Standard Model parameters.

After a relatively slow start due to large technical difficulties in the
mid-1990s, QCD calculations at next-to-next-to-leading order (NNLO) have
recently attained their goals for several phenomenological applications. The
aim of this Report is therefore to review recent progress in this field with
special emphasis on phenomenologically relevant results, including structure
functions, jet, and Higgs boson production. Very recently, twistor methods
have received particular attention, as they may offer a route to efficient
multiparticle calculations, even beyond tree-level. They will therefore be
briefly discussed, before we present our conclusions.

\section{Structure Functions}

Since its start-up in 1992, the DESY $ep$ collider HERA has provided a
wealth of data on the structure functions of the proton, and thus its parton
densities (PDFs), in a large region of Bjorken-$x$ and $Q^2$ \cite{Behnke}.
The PDFs represent at the same time an important ingredient in the
search for new physics, that is currently underway at the Fermilab Tevatron
and is an important research goal at the CERN LHC \cite{Roeck}. In
particular, QCD uncertainties in PDF determinations, {\it e.g.} from
renormalization and factorization scale and scheme \cite{Klasen:1996yk,%
Martin:2004ir} variations, have to be under control before disagreement
between theory and experiment can be interpreted as evidence for new
physics.

It is thus fortunate that, after completion of the three-loop singlet
splitting functions (obtained from the $1/\epsilon$-poles in dimensional
regularization) and longitudinal coefficient function (obtained from the
finite terms) \cite{Vogt:2004mw}, a
full NNLO calculation of the structure functions $F_{1,2}(x,Q^2)$ (and thus
also of $F_L = F_2 - 2 x F_1)$, is now available. These results have been
obtained using a large variety of techniques, such as application of the
optical theorem, transformation to Mellin space, mapping of diagrams with
composite topologies to basic building blocks, and integration by parts.
They also relied on recent advances in mathematics (harmonic sums) and
computer algebra (QGRAF, FORM). Fortunately, the results can also be applied
to other physical observables such as photon structure functions and total
cross sections in $e^+e^-$-annihilation.

\begin{figure}[h]
  \includegraphics[width=.29\textwidth]{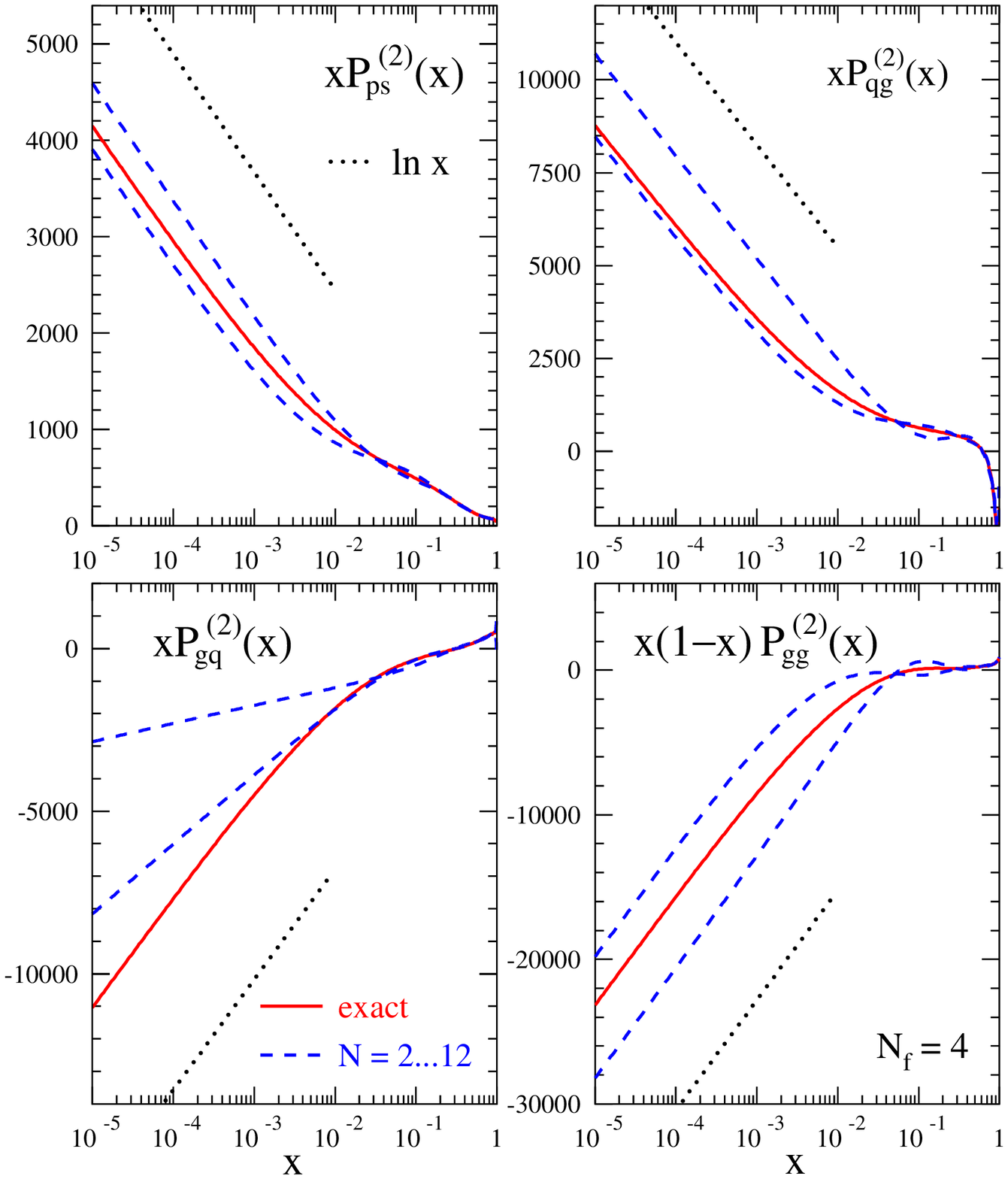}
  \includegraphics[width=.24\textwidth]{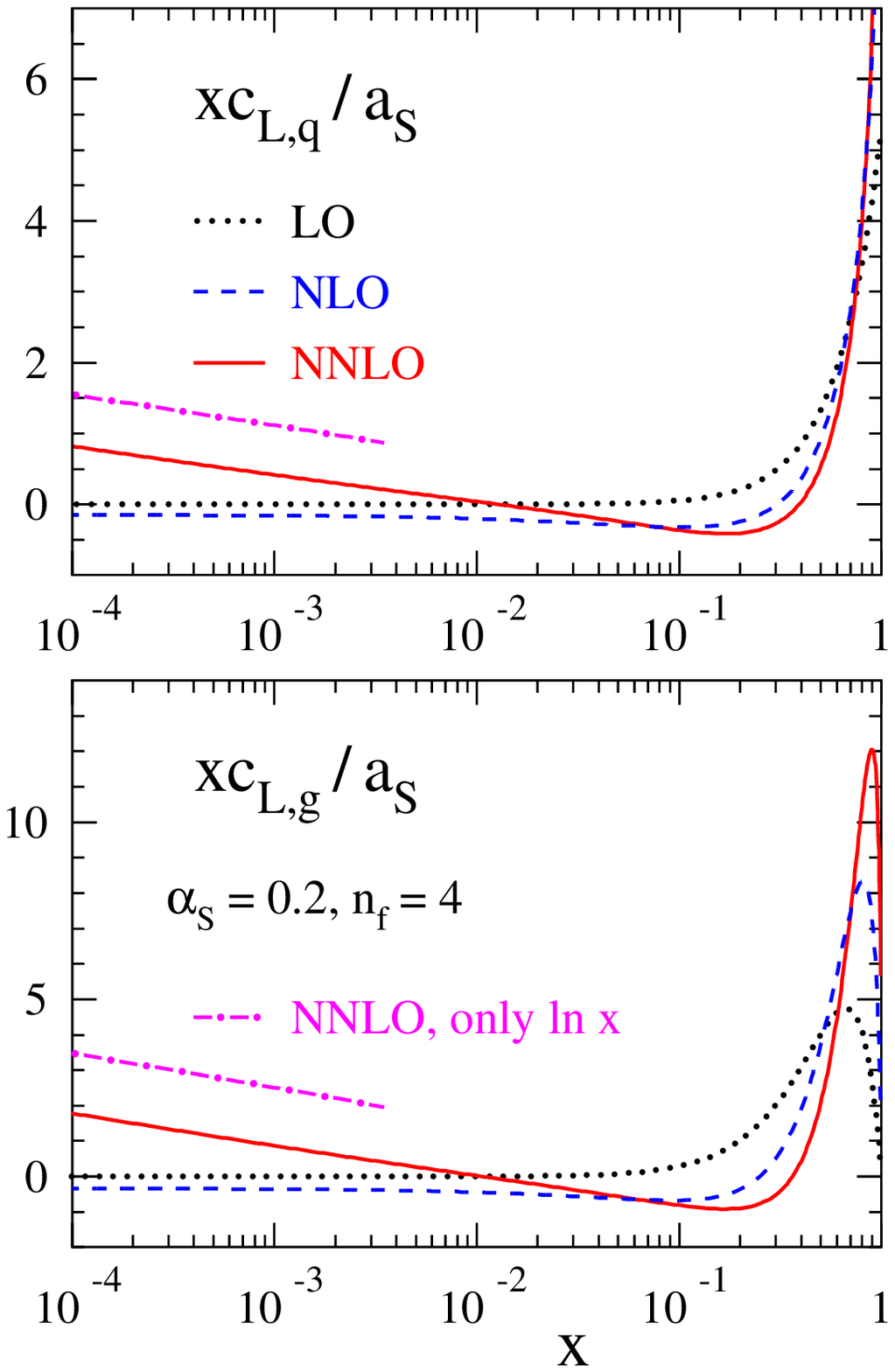}
  \caption{\label{fig:1}Splitting functions (left) and longitudinal
  coefficient functions (right) at NNLO \cite{Vogt:2004mw}.}
\end{figure}

While the numerical effects of the NNLO QCD corrections are well visible in
splitting functions and coefficient functions {\em separately} and show
clear differences in normalization, although not in shape, from earlier
leading-$\ln(x)$ estimates (see Fig.\ \ref{fig:1}), the
total effect in the {\em convoluted} structure functions is only
important for very small or large $x$ and small $Q^2$. For this reason, the
NNLO corrections have not yet been included in the global CTEQ analysis, as
the authors find sufficient stability of their fit at NLO
\cite{Huston:2005jm}, whereas the MRST collaboration have now implemented
the exact NNLO splitting functions, but find minor effects w.r.t.\ a
DIS-scheme motivated variation of the input gluon density
\cite{Martin:2004ir} or an earlier analysis using approximate NNLO splitting
functions \cite{Martin:2002dr}.

\section{Jet production in electron-positron annihilation}

Calculations of observables less inclusive than deep inelastic scattering
(DIS) structure functions and total $e^+e^-$ cross sections do not allow for
a straight-forward application of the optical theorem, as they require
multiparticle cuts in the three-loop photon propagator. For this reason,
NNLO jet calculations in $e^+e^-$-annihilation have been lagging behind,
although recently
considerable progress has also been achieved in this field. The interference
terms of $1\to3$ tree-level and two-loop amplitudes are now known, as are
the single soft and/or collinear regions of $1\to4$ squared one-loop
amplitudes. What remains to be calculated are the maximally double-soft and
triple-collinear regions of $1\to5$ squared tree-level amplitudes. Here,
antenna functions have successfully been used for quark-gluon and
gluon-gluon final states \cite{Gehrmann-DeRidder:2005hi}.

A first preliminary numerical result has been obtained for the NNLO
contribution to the $C_F^2$ color class of the average thrust
\cite{Gehrmann-DeRidder:2004xe}
\begin{eqnarray*}
 \langle 1-T \rangle &=& \int (1-T)\frac{1}{\sigma_0}\frac{d\sigma}{dT}
 ~~=~~ C_F \left[ \left(\frac{\alpha_s}{2\pi}\right) A +
 \left(\frac{\alpha_s}{2\pi}\right)^2 B+\left(\frac{\alpha_s}{2\pi}\right)^3
 C + \ldots\right],
\end{eqnarray*}
where $A=1.57$ and $B=32.3$ were known and now $C(C_F^2)=-20.4 \pm 4$.
The full result will be important for reliable estimates of higher-twist
effects in $e^+e^-$ event shapes. Furthermore, analytic continuation allows
to apply the results obtained for three-jet production in
$e^+e^-$-annihilation also to dijet production in DIS.

\section{Higgs boson production at hadron colliders}

\begin{figure}[h]
  \includegraphics[width=.40\textwidth]{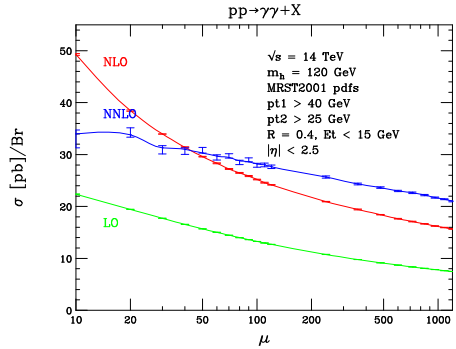}
  \caption{\label{fig:2}Higgs production at the LHC with diphoton decay and
  no additional hard jets \cite{Anastasiou:2004xq}.}
\end{figure}

The inclusive NNLO cross section for light/heavy scalar and pseudo-scalar
Higgs production in $gg$ scattering at hadron colliders has been known for
several years, but the relevance of these calculations has been limited by
the need for experimental cuts on associated photons, jets or $b$-quarks.
Fully differential cross sections, such as a recent NNLO calculation for
Higgs plus dijet production \cite{Anastasiou:2004xq},
allow for transverse energy vetos on the additional jets and thus for a
suppression of the QCD background in the diphoton (or other) Higgs decay
channels (see Fig.\ \ref{fig:2}).

\section{Jet hadroproduction and twistor methods}

For hadron-hadron scattering, the two-loop $2\to2$ parton amplitudes
have been known for some time, and the result for polarized $qq\to qq$
scattering has recently been confirmed using supersymmetric methods
\cite{DeFreitas:2004tk}. Since the soft/collinear singular regions of
one-loop $2\to3$ amplitudes are also known, recent work has focused on
multiple soft/collinear emissions in tree-level multiparton amplitudes and
on extending the dipole subtraction formalism to NNLO, where two
alternatives to previously known methods have been proposed
\cite{Frixione:2004is,Somogyi:2005xz}. Although the subtraction terms are in
principle known, the singular regions often overlap and need to be
disentangled to avoid double-counting. Furthermore, phase space
factorization and integration of the singular matrix elements still needs to
be resolved. For this reason, further progress towards a full NNLO
calculation for jet hadroproduction has been slow, and numerical integration
methods using sector decomposition \cite{Binoth:2004jv} may have to be
developped further.

Multiparticle amplitudes may be calculated efficiently by relating
perturbative gauge theory to the $D$-instanton expansion for a topological
string in twistor space. While originally motivated by
$N=4$ super Yang-Mills theory, this approach has been shown to work at
tree-level for generic QCD amplitudes. $N=4$ supersymmetric and
cut-constructible one-loop diagrams have also been calculated, and
extensions to non-supersymmetric theories and two-loop amplitudes are
underway \cite{Dixon}.

\section{Conclusion}

Although this brief Report is necessarily far from complete, it should be
clear that NNLO QCD calculations represent a theoretically challenging,
phenomenologically important, and fast-moving field of research allowing
for interesting advances in the fields of mathematics, computer science
and last, but not least, elementary particle physics.


\begin{theacknowledgments}
 The author thanks the working group convenors for the kind invitation, the
 DoE's INT at the University of Washington for kind hospitality and partial
 financial support during preparation of this work, and L.\ Dixon and T.\
 Teubner for useful discussions.
\end{theacknowledgments}



\end{document}